\documentclass[twocolumn,showpacs,amsmath,amssymb,nofootinbib]
{revtex4}
\usepackage{amsmath}
\usepackage{cancel}
\usepackage{physics}
\usepackage{subfigure}
\usepackage[T1]{fontenc}
\usepackage[utf8]{inputenc}
\usepackage{graphicx,dcolumn,bm,color,latexsym,amssymb}
\usepackage{import}
\usepackage{xifthen}
\usepackage{pdfpages}
\usepackage{transparent}
\usepackage{verbatim}
\usepackage{morefloats}
\usepackage{bm}
\newcommand{\beq}{\begin{equation}}
\newcommand{\eeq}{\end{equation}}

\usepackage[retainorgcmds]{IEEEtrantools}
\usepackage{graphicx,tikz,placeins}
\usepackage{mathrsfs}
\usepackage{amsmath,amssymb,amsfonts,physics}
\usepackage{color}
\usepackage{float}
\usepackage{times,txfonts}
\usepackage{nicefrac}
\usepackage{ulem}
\usepackage[colorlinks=true,linkcolor=blue,urlcolor=blue,citecolor=blue,pdfusetitle]{hyperref}
\usepackage{physics}
\usepackage{soul}
\usepackage{cancel} 
\usepackage{multirow}

\definecolor{Blue}{rgb}{0.0,0.0,1}
\definecolor{Red}{rgb}{1,0.0,0.0}
\definecolor{Green}{rgb}{0,0.5,0.0}

\begin{document}
\title{Statistics of power and efficiency for collisional Brownian engines}

\author{Gustavo A. L. For\~ao$^1$}

\author{Fernando S. Filho$^{1,2}$}

\author{Pedro V. Paraguass\'{u}$^3$}

\affiliation{$^1$Universidade de São Paulo,
Instituto de Física,
Rua do Matão, 1371, 05508-090
São Paulo, SP, Brazil}
\affiliation{$^2$UHasselt, Faculty of Sciences, Theory Lab, Agoralaan, 3590 Diepenbeek, Belgium}
\affiliation{$^3$Departamento de F\'{i}sica, Pontif\'{i}cia Universidade Cat\'{o}lica 22452-970, Rio de Janeiro, Brazil}

\date{\today}

\begin{abstract}
Collisional Brownian engines have attracted significant attention due to their simplicity, experimental accessibility, and amenability to exact analytical solutions. While previous research has predominantly focused on optimizing mean values of power and efficiency, the joint statistical properties of these performance metrics remain largely unexplored. Using stochastic thermodynamics, we investigate the joint probability distributions of power and efficiency for collisional Brownian engines, revealing how thermodynamic fluctuations influence the probability of observing values exceeding their respective mean maxima. Our conditional probability analysis demonstrates that when power fluctuates above its maximum mean value, the probability of achieving high efficiency increases substantially, suggesting fluctuation regimes where the classical power-efficiency trade-off can be probabilistically overcome. Notably, our framework extends to a broader class of engines, as the essential features of the statistics of the system are fully determined by the Onsager coefficients. Our results contribute to a deeper understanding of the role of fluctuations in Brownian engines, highlighting how stochastic behavior can enable performance beyond traditional thermodynamic bounds.

\end{abstract}

\maketitle

\section{Introduction}

In recent decades, the construction and characterization of nonequilibrium small engines have been broadly studied in the realm of stochastic thermodynamics \cite{seifert2008stochastic,tome2015stochastic,van2015ensemble}. On such scale, the fluctuations are unavoidable and play an important role in the system \cite{van2013stochastic,van2015ensemble,j2007statistical}. Among the relevant small-scale frameworks that deal with fluctuations within the realm of stochastic methods, we highlight molecular motors and rotors in cellular environments \cite{pietzonka2016universal,PhysRevLett.98.258102,busiello2022hyperaccurate,PhysRevE.92.042133,michl2009molecular,zimmermann2012efficiencies,seifert2011stochastic}, single active and inactive Brownian particles in thermal environment \cite{proesmans_brownian_2016,iago,PhysRevResearch.2.043262,abreu2011extracting,izumida2010onsager,schmiedl2007efficiency,dechant2017underdamped,berger2009optimal,gustavo,fernando,solon2015active,romanczuk2012active,
abdoli2021stochastic,speck2016stochastic,
speck2018active,speck2022critical}, quantum-dot pumps in open systems \cite{qd-carnot,harunari,lee,fernando2024thermodynamics,franklin}, non-equilibrium chemical reactions \cite{schmiedl2007stochastic,
rao2016nonequilibrium,
tome2018stochastic,
mou1986stochastic} and others.
Typically, such examples involve a non-equilibrium steady state (NESS) arising from either constant or periodic external forces, or from a  sequential setup where the system interacts with an uncorrelated thermal bath and a specific worksource at each stage. In the "collisional" approach, the system is modeled by assuming that the particle interacts sequentially with different, uncorrelated portions of the medium. In any given interval, it "collides" with one small fraction before moving on to another. Due to its simplicity and robustness, this approach has been applied across various scenarios and shown to be a viable alternative for efficient thermal engines with engineered reservoirs, including Brownian particle systems operating as work-to-work converters \cite{gustavo,fernando}, quantum dot pumps functioning as heat engines \cite{harunari,fernando2024thermodynamics,franklin,lee}, or even minimal interacting systems
\cite{collective3}. In particular, the collisional approach has also been widely used in open quantum systems \cite{giovannetti2012master, franklin} and to study memory degradation in quantum and classical systems \cite{parrondo}.

A key difference between equilibrium and nonequilibrium
statistical mechanics is that the latter allows for obtaining statistical distributions of quantities such as heat flux, power, and entropy production, rather than just their average values. This reveals additional features, such as the existence of "forbidden" values, including negative entropy production or efficiencies greater than Carnot \cite{crooks1999entropy,paraguassu2022probabilities, barros2024probabilistic,salazar2021detailed,polettini_efficiency_2015}, which, although rare, can still occur. It also opens the possibility of measuring values greater than the optimized mean quantities for power and efficiency. While this approach has been applied in a variety of classical \cite{verley2014universal,proesmans2016brownian,holubec_fluctuations_2021,proesmans_stochastic_2015,proesmans_stochastic_2015-1, holubec_efficiency_2015, park_efficiency_2016, vroylandt_efficiency_2020} and quantum systems \cite{barra_efficiency_2022, esposito_efficiency_2015, denzler_efficiency_2020,denzler_power_2021, fei_efficiency_2022}, there are still open
questions about measuring of rare events in collisional frameworks and such a scenario is far from
being fully understood. 

In this contribution, we address this gap by studying rare event statistics in a well-known collisional work-to-work converter, where a single Brownian particle is placed sequentially with two thermal
reservoir with different time-dependent external drives \cite{fernando,gustavo,noa2021efficient}. While related studies have examined efficiency distributions \cite{polettini_efficiency_2015,proesmans_stochastic_2015}, our work focuses on collisional Brownian systems and investigates regimes in which power and efficiency surpass their mean-optimal values, along with a detailed analysis of their joint and conditional distributions.
In doing so, we extend the scope of previous investigations \cite{verley2014universal,proesmans2016brownian,holubec_fluctuations_2021,proesmans_stochastic_2015,proesmans_stochastic_2015-1, holubec_efficiency_2015, park_efficiency_2016, vroylandt_efficiency_2020, barra_efficiency_2022, esposito_efficiency_2015, denzler_efficiency_2020,denzler_power_2021, fei_efficiency_2022,polettini_efficiency_2015} by examining the full probability landscape of performance, including both marginal and joint distributions of power and efficiency. This enables a deeper investigation into their statistical behavior and the chances of specific events, allowing us to ask questions such as: "What is the probability of the efficiency falling within a certain range, given that the power already meets a specific value?". Furthermore, we examine the full distribution of these quantities, both conditionally and unconditionally, to understand how they influence one another. This analysis sheds light on the rare but significant occurrences of\,\,"forbidden" events and explores the possibility of measuring values greater than the optimized mean quantities of power and efficiency. Overall, this framework provides a more comprehensive understanding of the complex interplay between power, efficiency, entropy production, and fluctuations in nonequilibrium thermodynamics.

This paper is organized as follows: In Section \ref{II} we introduce the general model and thermodynamics for a single cyclic Brownian engine.  The optimization protocol in terms of the external driving for the mean values of the output performance is summarized in the Section \ref{IV}. The introduction of the probability distribution for thermodynamical quantities and its joint case are presented in the Section \ref{III}. In Section \ref{V}, we explore the probability distribution of efficiency and power and examine how to optimize both quantities for superperformance values. In Section \ref{VI}, we derive the conditional distribution of efficiency, given that the power lies within a specific range, and explore the implications of this condition by analyzing two distinct scenarios. We finish in Section \ref{conclusion} with the discussion and conclusions of the results.

\section{Thermodynamics of collisional Brownian engines}
\label{II}

The proposed system consists of a Brownian particle of mass m that alternates between two thermal reservoirs while being subjected to an external driving force, $F_i(t)$, during each stroke $i$ of the process, where $i \in \{1,2\}$. In the first stroke, the particle is in contact with the reservoir at temperature $T_1$ and experiences the force $F_1(t)$ for a duration $\tau/2$. At $t = \tau/2$, it transitions to a second thermal bath at temperature $T_2$, where it is subjected to the force $F_2(t)$ for the same time interval. When $t = \tau$, the cycle completes and starts again. Three points are worth highlighting. First, the two-stroke framework is chosen for its analytical tractability and established experimental relevance \cite{proesmans2016brownian}, enabling an analysis that is simultaneously simple and rich in physical insights. Second, we assume equal stroke durations for simplicity, allowing a clear focus on our main goal, i.e., the investigation of novel optimization strategies arising from the statistical nature of the system. We note, however, that asymmetric durations could also be explored, as it has been shown that it can improve mean performance \cite{noa2021efficient,harunari}. Third, in the collisional model, the exchange between reservoirs is assumed to be instantaneous, meaning no heat is exchanged during the transition. The dynamics of the Brownian particle is described by the Langevin equation

\begin{equation}
    \frac{d v_i}{dt} =  - \gamma_i v_i(t) + F_i(t)+\xi_{i}(t),
  \label{eq3}
\end{equation}

\noindent where $v_i$ is the velocity of the particle at stroke $i$, $\gamma_i$ represents the viscous coefficient and $\xi_i(t)$ is the stochastic force following the standard white noise properties $\langle \xi_i(t)\rangle = 0$ and $\langle\xi_i(t)\xi_j(t')\rangle = 2\,\gamma_i\,k_b\,T_i\,\delta_{ij}\,\delta(t-t')/m$. The driving force (per mass) consists of two components, each corresponding to one of the cycle's strokes, namely

    \begin{equation}
    F_i(t) =
    \begin{cases}
    X_1\,g_1(t), & 0 \leq t < \tau/2, \\
    X_2\,g_2(t), & \tau/2 \leq t < \tau,
    \end{cases}
    \end{equation}

\noindent with \(X_i\) denoting the strength of the thermodynamic force applied in each stage and \(g_i(t)\) defining the protocol shape (e.g. constant, linear or sinusoidal) over the corresponding interval. 
This approach offers a simplified model of Langevin-like procedures for describing nanoscale transport phenomena, such as those found in the cellular realm \cite{lopez2008realization,prost1994asymmetric,astumian1994fluctuation}. Specifically, the selection of forces $F_i$
for each stroke is analogous to applying a time-dependent, asymmetric potential to a diffusing particle in a ratchet system, which can be achieved with chemical potentials in active systems or optical traps in passive systems \cite{lopez2008realization}. This setup effectively models, for example, the movement of a single kinesin molecule along a biopolymer.
By averaging Eq.~(\ref{eq3}) together the boundary conditions ensure that $\langle v_1\rangle(\tau/2)=\langle v_2\rangle(\tau/2)$
and $\langle v_1\rangle(0)=\langle v_2\rangle(\tau)$. The mean velocity $\langle v_i\rangle$ for the $i$-th
stage is thus given by
\begin{widetext}
        \begin{eqnarray}
        \langle v_1(t)\rangle = X_1 \left(\frac{ e^{-\frac{\gamma  t}{m}} \int _0^{\frac{\tau }{2}}e^{\frac{t' \gamma }{m}} g_1(t')dt'}{m \left(e^{\frac{\gamma  \tau }{m}}-1\right)}+\frac{ e^{-\frac{\gamma  t}{m}} \int _0^te^{\frac{t' \gamma }{m}} g_1(t')dt'}{m}\right) + X_2 \left(\frac{ e^{-\frac{\gamma  t}{m}} \int _0^{\tau }e^{\frac{t' \gamma }{m}} g_2(t')dt'}{m \left(e^{\frac{\gamma  \tau }{m}}-1\right)}-\frac{ e^{-\frac{\gamma  t}{m}} \int _0^{\frac{\tau }{2}}e^{\frac{t' \gamma }{m}} g_2(t')dt'}{m \left(e^{\frac{\gamma  \tau }{m}}-1\right)}\right) \label{v1}, \\
       \langle v_2(t)\rangle = X_1 \left(\frac{ e^{\frac{\gamma  (\tau -t)}{m}} \int _0^{\frac{\tau }{2}}e^{\frac{\gamma  t'}{m}} g_1(t')dt'}{m \left(e^{\frac{\gamma  \tau }{m}}-1\right)}\right) + X_2 \left(-\frac{e^{\frac{\gamma  (\tau -t)}{m}} \int _0^{\frac{\tau }{2}}e^{\frac{\gamma  t'}{m}} g_2(t')dt'}{m \left(e^{\frac{\gamma  \tau }{m}}-1\right)}+\frac{e^{-\frac{\gamma  t}{m}} \int _0^{\tau }e^{\frac{\gamma  t'}{m}} g_2(t')dt'}{m \left(e^{\frac{\gamma  \tau }{m}}-1\right)}+\frac{e^{-\frac{\gamma  t}{m}} \int _0^te^{\frac{\gamma  t'}{m}} g_2(t')dt'}{m}\right)\label{v2},
    \end{eqnarray}
    \end{widetext}
    The model thermodynamics can be setup by examining the time-dependent probability distribution \( P_i(v, t) \), described by the following Fokker-Planck equation.

\begin{equation}
\label{FK}
    \frac{\partial P_i}{\partial t} = -\left[ {F}_i(t)\,\frac{\partial P_i}{\partial v}+\frac{\partial J_i}{\partial v} \right],
\end{equation}

\noindent where $J_i$ is a current of probability given by $J_i = -\gamma_i\,v\,P_i - \frac{\gamma_i\,k_B\,T_i}{m}\,\frac{\partial P_i}{\partial v}$. A key aspect of Eq.~(\ref{FK}) is noteworthy: the probability distribution maintains a Gaussian shape irrespective of the applied temperatures or external forces \cite{angel,fernando}, which is a crucial characteristic for deriving the system's thermodynamic statistical properties in Section \ref{III}. By taking the time evolution of the mean energy $U_i(t) = m  \langle v_i^2(t)\rangle /2$, one finds that it can be expressed as the sum of two contributions, $dU_i(t)/dt = -\left[\dot{W}_i(t) +\dot{Q}_i(t)\right]$, where the first and second terms on the right represent the work per unit time done on (extracted from) the particle by the force $F_i(t)$ and the heat flux $\dot{Q}_i(t)$ to (from) the thermal bath during stroke $i$. Specifically,
\begin{equation}
\label{work}
    {\dot{W}}_i(t) = -X_i\,g_i(t)\, \langle v_i(t) \rangle,
\end{equation}
\begin{equation}
    \dot{Q}_i(t) = \gamma_i\left(m\, \langle v_i^2 (t)\rangle - k_B\,T_i\right).
\end{equation}
\newline
The second law of thermodynamics is related to the time evolution of entropy \( S_i = -k_B \langle \ln{P_i} \rangle \). This relationship, in conjunction with Eq.~(\ref{FK}), can be described by the difference between the entropy production rate \( \Sigma_i(t) \) and the entropy flux \( \Phi_i(t) \), $
    dS_i/dt = \Sigma_i(t) - \Phi_i(t),
$
where
\begin{equation}
\label{epeq}
   \Sigma_i(t) = \frac{m}{\gamma_i\,T_i}\int \frac{J_i^2}{P_i}\,dv\,\,\,\,\,\,\,\,\,\,\textrm{and}\,\,\,\,\,\,\,\,\,\, \Phi_i(t) = \frac{\Dot{Q}_i(t)}{T_i}.
\end{equation}
\newline
\noindent It is evident that \(\Sigma_i(t) \geq 0\), consistent with the second law of Thermodynamics. By averaging it over a complete cycle, one has
\begin{align}
    \,\,\,\,\,{\Sigma}& = \frac{1}{\tau} \left( \int_{0}^{\tau/2} d\Sigma_1(t) \, dt + \int_{\tau/2}^{\tau} d\Sigma_2(t) \, dt \right) = 
    \\ \label{cep}&= \frac{4 T^2}{4 T^2 - \Delta T^2} \left[ \frac{-\left({\bar{\dot{W}}}_1 + {\bar{\dot{W}}}_2\right)}{T} + \frac{\left({\bar{\dot{Q}}}_1 - {\bar{\dot{Q}}}_2\right) \Delta T}{2 T^2} \right],
\end{align}
\noindent
where  $\bar{\dot{W}}_1=(1/\tau)\int_{0}^{\tau/2}  {\dot{W}}_1(t) \, dt$ and $\bar{\dot{W}}_2=(1/\tau)\int_{\tau/2}^{\tau}  {\dot{W}}_2(t) \, dt$, respectively, and
\(T = (T_1 + T_2)/2\) and \(\Delta T = T_1 - T_2\) are introduced together the first law of thermodynamics. The evaluation of the above quantities over the corresponding state allows
to write it in the following form
\begin{equation}
    \bar{\dot{W}}_i =  -T (L_{ii}f_i^{2} + L_{ij} f_i f_j),
    \label{w11}
\end{equation}
where $f_i=X_i/T$ accounts for the thermodynamic force and the dependence of $\bar{\dot{W}}_i$ on $f_j$ arises from the cyclic boundary conditions that couple the two strokes. The $L_{ij}$'s are Onsager coefficients given by 
\begin{widetext}
    \begin{eqnarray}
        L_{11}= -\int_{0}^{\tau/2} \int_{0}^{t} \frac{e^{\frac{s - t \, \gamma}{m}} T g_1(s) g_1(t)}{(-1 + e^{\frac{\gamma \tau}{m}}) m \tau} \, ds \, dt 
+ \int_{0}^{\tau/2} \int_{0}^{t} \frac{e^{\frac{s - t \, \gamma}{m} + \frac{\gamma \tau}{m}} T g_1(s) g_1(t)}{(-1 + e^{\frac{\gamma \tau}{m}}) m \tau} \, ds \, dt 
+ \int_{0}^{\tau/2} \int_{0}^{\tau/2} \frac{e^{\frac{s - t \, \gamma}{m}} T g_1(s) g_1(t)}{(-1 + e^{\frac{\gamma \tau}{m}}) m \tau} \, dt \, ds, \\
L_{12} = -\int_{0}^{\tau/2} \int_{0}^{\tau/2} \frac{e^{\frac{s - t \, \gamma}{m}} T g_1(t) g_2(s)}{(-1 + e^{\frac{\gamma \tau}{m}}) m \tau} \, dt \, ds 
+ \int_{0}^{\tau} \int_{0}^{\tau/2} \frac{e^{\frac{s - t \, \gamma}{m}} T g_1(t) g_2(s)}{(-1 + e^{\frac{\gamma \tau}{m}}) m \tau} \, dt \, ds, \\
L_{21} = \int_{0}^{\tau/2} \int_{\tau/2}^{\tau} \frac{e^{\frac{\gamma (s - t + \tau)}{m}} T g_1(s) g_2(t)}{(-1 + e^{\frac{\gamma \tau}{m}}) m \tau} \, dt \, ds, \\
L_{22} = 
\int_{\tau/2}^{\tau} \int_{0}^{t} \left(-\frac{e^{\frac{s - t \, \gamma}{m}} T g_2(s) g_2(t)}{(-1 + e^{\frac{\gamma \tau}{m}}) m \tau}\right) \, ds \, dt 
+ \int_{\tau/2}^{\tau} \int_{0}^{t} \frac{e^{\frac{s - t \, \gamma}{m} + \frac{\gamma \tau}{m}} T g_2(s) g_2(t)}{(-1 + e^{\frac{\gamma \tau}{m}}) m \tau} \, ds \, dt 
+ \int_{\tau/2}^{\tau} \int_{0}^{\tau/2} \left(-\frac{e^{\frac{s - t \, \gamma}{m} + \frac{\gamma \tau}{m}} T g_2(s) g_2(t)}{(-1 + e^{\frac{\gamma \tau}{m}}) m \tau}\right) \, dt \, ds 
\nonumber\\+ \int_{\tau/2}^{\tau} \int_{0}^{\tau} \frac{e^{\frac{s - t \, \gamma}{m}} T g_2(s) g_2(t)}{(-1 + e^{\frac{\gamma \tau}{m}}) m \tau} \, dt \, ds.\label{onsager}
    \end{eqnarray}
\end{widetext}
Expressions similar to (\ref{w11}) hold
for the ${\bar{\dot{Q}}}_i$'s. 
By properly adjusting \(X_1\) and \(X_2\) during one of the strokes, such class of engines operates as a work-to-work converter,  which is a group of engines that transform one kind of work into another, dissipating heat during the process. Such a framework is prominent in experimental designs \cite{proesmans2016brownian} and also present in biological examples, such as molecular motors, which are enzymes capable of transforming chemical work into mechanical work \cite{seifert2008stochastic,seifert2011stochastic,busiello2022hyperaccurate}.
For completeness, we highlight that such a system cannot work as a heat engine, since its framework enables it to transduce only one 
mechanical flux into another. Therefore, the temperature gradient between the two reservoirs does not play any relevant role in optimizing the work output and, in fact, decreases the efficiency, since the heat fluxes increase dissipation without contributing to the output work \cite{fernando}.

For this reason, we shall focus on the simplest case
$\Delta T=0 \rightarrow T_1 = T_2 = T$ in such a way that Eq.~(\ref{cep}) simplifies to:
\begin{equation}
\label{isothermal_entrop}
    {\Sigma} =-\frac{\left({\bar{\dot{W}}}_1 + {\bar{\dot{W}}}_2\right)}{T}= L_{11}f_1^{2} + (L_{12}+L_{21}) f_1 f_2+L_{22}f_2^2.
\end{equation}

In the case of a work-to-work converter,   the input work rate \({\bar{\dot{W}}}_{\text{in}} = {\bar{\dot{W}}}_i<0\) is partially converted into output work ${\bar{\dot{W}}}_{\text{out}} = {\bar{\dot{W}}}_j\geq 0$. The efficiency of this conversion is defined as
\begin{equation}
\overline\eta \equiv -\frac{{\bar{\dot{W}}}_{\text{out}}}{{\bar{\dot{W}}}_{\text{in}}},\label{avgeff}
\end{equation}
where  $0 \leq \overline\eta \leq 1$
 and solely depend on Onsager coefficients
and thermodynamic forces. Note that $\overline{\eta}$ generally differs from the stochastic efficiency $\eta$, which will be formally defined in Section \ref{III}. Here, the power and efficiency introduced correspond to the mean values of stochastic power and stochastic entropy production, whereas in Section \ref{III}, they can take any value fluctuating around the mean. Before going to the fluctuations, we first revisit the optimization of the mean power and efficiency, as discussed in \cite{karelonsager,fernando}.

\section{Overview about mean power and efficiency optimizations}
\label{IV}

The mean values of power and efficiency can be optimized based on the output power strength. Different protocols and external force profiles have been previously analyzed in \cite{fernando}. Here, we briefly review these results as a reference for our probabilistic analysis. By fixing \( f_1 \), we determine the value of \( f_2 \) that maximizes the mean power and the value that maximizes the mean efficiency, which are, respectively,

\begin{eqnarray}
    f_2^{\rm MP }  = - \frac{1}{2} \frac{L_{21}}{L_{22}} f_1, \label{maxpower}\\f_2^{\rm ME } = \frac{L_{11}}{L_{12}}\left(-1+\sqrt{1-\frac{L_{12}L_{21}}{L_{11}L_{22}}}\right)f_1.\label{maxeff}
\end{eqnarray}

Notice that the optimized thermodynamic forces depend only on the Onsager coefficients, which ensures that these results hold for arbitrary protocols. Using these values, one can calculate the mean maximum power, i.e., $\bar{\dot{W}}_2^{\rm MP} = \bar{\dot{W}}_2(f_2^{\rm MP})$, and the mean power at maximum efficiency, $\bar{\dot{W}}_2^{\rm ME} = \bar{\dot{W}}_2(f_2^{\rm ME})$. The expression for the maximum efficiency and the efficiency at maximum power can be calculated analytically and are respectively given by

\begin{equation}
    \bar{\eta}_{\rm ME} = -\frac{L_{21}}{L_{12}} + \frac{2L_{11}L_{22}}{L_{12}^2}\left(1-\sqrt{1-\frac{L_{12}L_{21}}{L_{11}L_{22}}}\right).\label{maxeffi}
\end{equation}
\begin{equation}
    \bar{\eta}_{\rm MP} = \frac{L_{21}^2}{4 L_{11}L_{22}-2L_{12}L_{21}}, 
\end{equation}

\begin{figure}
    \centering
    \includegraphics[width=8.6cm]{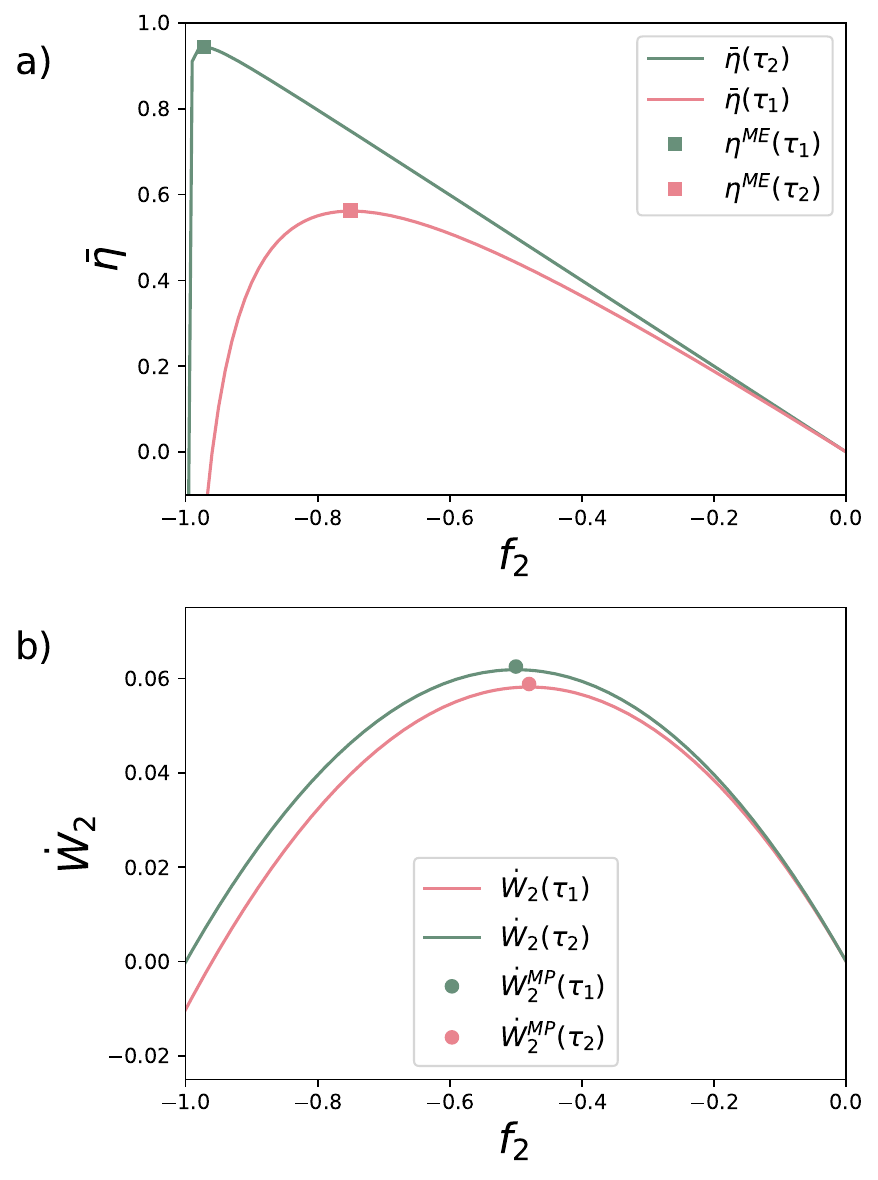}
    \caption{Depiction of the efficiency ${\bar \eta}$ (top) 
    and mean power $\bar{\dot{W}}_2$ (bottom) versus different $X_2=Tf_2$
    for different $\tau$'s. Green curves show results for $\tau_1=0.1$, with 
    $f_2^{\rm ME}=-0.972$, $\eta^{\rm ME}=0.944$, $\dot{W}_2^{\rm ME}=0.0068$ at maximum efficiency, and
    $f_2^{\rm MP}=-0.500$, $\eta^{\rm MP}=0.499$, $\dot{W}_2^{\rm MP}=0.0625$ at maximum power.
    Pink curves show results for $\tau_2=1$, with
    $f_2^{\rm ME}=-0.750$, $\eta^{\rm ME}=0.563$, $\dot{W}_2^{\rm ME}=0.0401$ at maximum efficiency, and
    $f_2^{\rm MP}=-0.480$, $\eta^{\rm MP}=0.428$, $\dot{W}_2^{\rm MP}=0.0588$ at maximum power.
    Symbols denote the corresponding maximum values $\bar{\dot{W}}_2^{MP}$
    and $\eta_{ME}$.}
    \label{overview}
\end{figure}

Having established these general results, we now turn our attention to a specific case that will serve as the foundation for the analysis that follows. To focus on the fundamental behavior of the system, without loss of generality, we restrict ourselves to the case of constant drivings, 
$g_1(t) = g_2(t) = 1$. This choice, already explored in \cite{fernando}, allows us to set a reference framework for our investigation. Working in reduced units where $m = T = \gamma = 1$, we take $f_1 = 1$ as a fixed parameter, which in turn constrains $f_2$ to the range $[-f_1, 0]$ to ensure a proper work-to-work conversion. As an example of the system's behavior, Fig.~\ref{overview} shows the results for two different periods, $\tau_1 = 0.1$ and $\tau_2 = 1$. One can observe that while the influence of the period on (mean) power and its optimal values remains mild, its impact on efficiency is more pronounced, with decreasing $\tau$ leading to an increase in $\bar{\eta}$, particularly at $\bar{\eta}_{\rm ME}$ and $f_2^{\rm ME}$.

\section{Probability description of power and efficiency}
\label{III}

Up to this point, all analyses have focused on the mean values. We now extend our discussion to the statistics of power and efficiency, providing a more comprehensive understanding of the system. Starting with the power distribution, we consider the power $\bar{\dot w}_i[x]$ along a stochastic trajectory given by
\begin{eqnarray}
    \bar{\dot w}_1[x] &=& - \frac{1}{\tau} \int_0^{\tau/2} F_1(t)\,v(t)\,dt,\nonumber\\ \bar{\dot w}_2[x] &=& - \frac{1}{\tau} \int_{\tau/2}^{\tau} F_2(t)\,v(t)\,dt,
\end{eqnarray}
where $F_{1}(t) = 0$, $(F_{2}(t)=0)$ at $\tau/2\le t<\tau~,(0< t\le\tau/2)$.
Note that  by averaring them, we promptly recover the
expressions for $\bar{\dot{W}}_i$'s (below exemplified for $\bar{\dot{W}}_1$)
\begin{eqnarray}
    \langle \bar{\dot w}_1[x]\rangle &=& - \frac{1}{\tau}\int_0^{\tau/2} F_1(t)\,\langle v(t)\rangle\,dt \nonumber\\ &=& - \frac{1}{\tau}\int_0^{\tau/2} F_1(t)\, \langle v_1(t)\rangle\,dt \nonumber\\ &=& \bar{\dot W}_1.
\end{eqnarray}

The above expression shows that $\bar{\dot w}_i[x]$ is linearly proportional to the stochastic velocity \( v(t) \), which follows a Gaussian probability distribution. Consequently, both $\bar{\dot w}_1$ and $\bar{\dot w}_2$ also exhibit a Gaussian form, with their correlation arising from the boundary conditions. Their joint distribution, $P_G(\bar{\dot w}_1,\bar{\dot w}_2)$, where the subscript $G$ explicitly indicates its Gaussian nature, is given by

\begin{equation}
    P_G(\bar{\dot w}_1,\bar{\dot w}_2) = \frac{1}{\sqrt{\det(C)}} \exp\left( - \frac{1}{2}\sum_{i,j=1}^2 \left(\bar{\dot w}_i - {\bar{\dot  W}_i}\right)C_{ij}^{-1}\left(\bar{\dot w}_j - \bar{\dot W}_j\right)\right),
\end{equation}
expressed in terms of averages $\bar{\dot W}_i = \langle \bar{\dot w}_i\rangle$ obeying Eq.~(\ref{w11}). The covariance matrix $C$ is defined by
\begin{equation}
    C_{ij} = \langle \bar{\dot w}_i\bar{\dot w}_j\rangle  - \langle \bar{\dot w}_i\rangle \langle \bar{\dot w}_j \rangle.
\end{equation}
Analogous to the mean values, covariances can also be expressed in terms of the Onsager coefficients. Following the work of Ref. \cite{proesmans2016brownian}, we observe that the stochastic work is related to the stochastic entropy fluxes by
\begin{equation}
    \sigma_1 = -\frac{\bar{\dot w}_1}{T},\; \sigma_2 = -\frac{\bar{\dot w}_2}{T}.
\end{equation}
Since the total entropy production follows a fluctuation theorem in the asymptotic limit, the joint distribution of the entropy fluxes must satisfy \cite{proesmans2016brownian}
\begin{equation}
    \frac{P(\sigma_1,\sigma_2)}{P^\dagger(-\sigma_1, -\sigma_2)} = \exp\left[\left(\sigma_1+\sigma_2\right)\tau/k_B\right],
\end{equation}
where $P^{\dagger}$ is the distribution of the reversed process. Given that the joint distribution is Gaussian, the fluctuation theorem can be used to determine the covariance matrix. Furthermore, since the powers are linearly related with the entropy production, the fluctuation theorem also applies to them, allowing us to obtain their covariance matrix following the same approach as in \cite{proesmans2016brownian}. As a result, we express the covariance components of the powers in terms of the Onsager coefficients.
\begin{equation}
    C_{ij} = \frac{T^2k_B}{\tau} \left(L_{ij}+L_{ji}\right).\label{covariance}
\end{equation}
At this point, it is important to highlight that no assumptions beyond the white noise Langevin equation description of the system have been made. Since the power statistics are entirely determined by the Onsager coefficients, our approach is applicable to a broader class of collisional systems beyond the one studied here. We conclude by noting that, given the joint distribution, the marginal distributions can be obtained simply by integration, i.e.,
\begin{equation}
    P(\bar{\dot w}_2) = \int_{-\infty}^{\infty} P(\bar{\dot w}_1, \bar{\dot w}_2)\,d\bar{\dot w}_1.
\end{equation}

Building on our analysis of power distributions, we now examine the stochastic efficiency, which incorporates the full correlations between input and output powers that the mean description in Eq. (\ref{avgeff}) neglects. We define it as
\begin{equation}
    \eta = - \frac{\bar{\dot w}_2}{\bar{\dot w}_1}.
\end{equation}
We highlight that, since this quantity represents the ratio of two random Gaussian variables, the efficiency distribution
\begin{equation}
    P(\eta) = \int_{-\infty}^{\infty}\int_{-\infty}^{\infty} P(\bar{\dot w}_1,\bar{\dot w}_2)\,\delta\left(\eta + \frac{\bar{\dot w}_2}{\bar{\dot w}_1}\right)\,d\bar{\dot w}_1\,d\bar{\dot w}_2.\label{effdef}
\end{equation}
\noindent admits an analytical solution \cite{proesmans2016brownian, proesmans_stochastic_2015-1}. To evaluate this integral, we first rewrite the Dirac delta function as
\begin{equation}
    \delta\left(\eta + \frac{\bar{\dot w}_2}{\bar{\dot w}_1}\right) = |\bar{\dot w}_1|\,\delta(\eta\,\bar{\dot w}_1 + \bar{\dot w}_2) ,
\end{equation}
allowing us to evaluate one integral directly. The remaining Gaussian integral can be solved analytically, yielding the result:
\begin{widetext}
    \begin{eqnarray}
        P(\eta) = \frac{\sqrt{-C_{12}^2 + C_{11} C_{22}} \, }{   \pi \left(C_{22} + \eta (2 C_{12} + C_{11} \eta)\right)} \exp\left({\frac{C_{22} \bar{\dot{W}}_1^2 - 2 C_{12} \bar{\dot{W}}_1 \bar{\dot{W}}_2 + C_{11} \bar{\dot{W}}_2^2}{2 C_{12}^2 - 2 C_{11} C_{22}}}\right)
+ \\ \frac{
    e^{-\frac{(\bar{\dot{W}}_2 + \bar{\dot{W}}_1 \eta)^2}{2 \left(C_{22} + \eta (2 C_{12} + C_{11} \eta)\right)}} 
    \left(C_{22} \bar{\dot{W}}_1 - C_{12} \bar{\dot{W}}_2 + C_{12} \bar{\dot{W}}_1 \eta - C_{11} \bar{\dot{W}}_2 \eta\right) 
    }{  \sqrt{2\pi} \left(C_{22} + \eta (2 C_{12} + C_{11} \eta)\right)^{3/2}}\text{Erf}\left[\frac{C_{22} \bar{\dot{W}}_1 - C_{12} \bar{\dot{W}}_2 + C_{12} \bar{\dot{W}}_1 \eta - C_{11} \bar{\dot{W}}_2 \eta}{\sqrt{2} \sqrt{\left(-C_{12}^2 + C_{11} C_{22}\right) \left(C_{22} + \eta (2 C_{12} + C_{11} \eta)\right)}}\right]
    \end{eqnarray}
\end{widetext}

Note that this distribution exhibits Cauchy-like behavior ($P(\eta) \sim 1/\eta^2$), resulting in ill-defined moments \cite{feller1950}, which has been identified as a universal property of efficiency distributions \cite{proesmans_stochastic_2015}. Moreover, it depends solely on the statistical moments of the power variables, with detailed analyses available in \cite{gupta_stochastic_2017, verley2014universal}. Crucially, in the next section, we employ it to calculate probabilities for exceeding the optimal efficiency in Eq.(\ref{maxeffi}).

Having derived both the power and efficiency distributions, we now examine their interrelationship. The intrinsic efficiency-power trade-off \cite{gustavo,fernando} motivates analyzing their joint statistics through an unified distribution, since both quantities originate from the same particle trajectory dynamics. This joint statistical framework enables conditional probability analysis while simultaneously capturing their fundamental thermodynamic competition. The joint distribution, $P(\eta,\bar{\dot w}_2)$, can be derived by noticing that the integrand in Eq.(\ref{effdef}) corresponds to
\begin{equation}
    P(\eta,\bar{\dot w}_1, \bar{\dot w}_2 ) = P_G(\bar{\dot w}_1,\bar{\dot w}_2)\,\delta\left(\eta + \frac{\bar{\dot w}_2}{\bar{\dot w}_1}\right).
\end{equation}

Therefore, to obtain the marginal $P(\eta,\bar{\dot w}_2)$, we need to integrate over $\bar{\dot w}_1$, which is simplified through the identity

\begin{equation}
    \delta \left(\eta + \frac{\bar{\dot w}_2}{\bar{\dot w}_1}\right) = \frac{|\bar{\dot w}_1|}{|\eta|} \delta \left(\bar{\dot w}_1 + \frac{\bar{\dot w}_2}{\eta}\right).
\end{equation}

Applying this transformation and performing the $\bar{\dot w}_1$ integration yields

\begin{equation}
    P(\eta,\bar{\dot w}_2) = \frac{|\bar{\dot w}_2|}{\eta^2} P_G\left(-\frac{\bar{\dot w}_2}{\eta},\bar{\dot w}_2\right).\label{jointdist}
\end{equation}
This distribution exhibits distinct behavior in each variable: it remains Gaussian in $\bar{\dot{w}}_2$ while preserving the characteristic Cauchy-like form in $\eta$. This complete statistical framework provides the foundation for analyzing the coupled dynamics between efficiency and power, which we develop in the following section.

\section{Optimization Beyond Mean Values}
\label{V}

Having established the probability distributions and the maximum values for the mean power and mean efficiency, we now turn our attention to computing the probabilities of specific statistical events. Due to fluctuations inherent in the stochastic power and efficiency, it is possible to observe values exceeding their respective mean maximum values. These probabilities can be determined by integrating the corresponding probability distributions. Our objective is to identify the values of the thermodynamic force $f_2$ that maximize the probability of such events occurring. 
\subsection{Power optimization}
We begin by analyzing the probability of power surpassing its maximum mean value. The average power, $\bar{\dot{W}}_2$, is constrained by an upper bound, such that $\bar{\dot{W}}_2 \leq \bar{\dot{W}}_2^{MP}$. However, because the stochastic power follows a Gaussian distribution, it is possible to derive an analytical expression for the probability of observing power values greater than $\dot W_2^{MP}$:

\begin{figure}  
    \centering  
    \includegraphics[width=8.6cm]{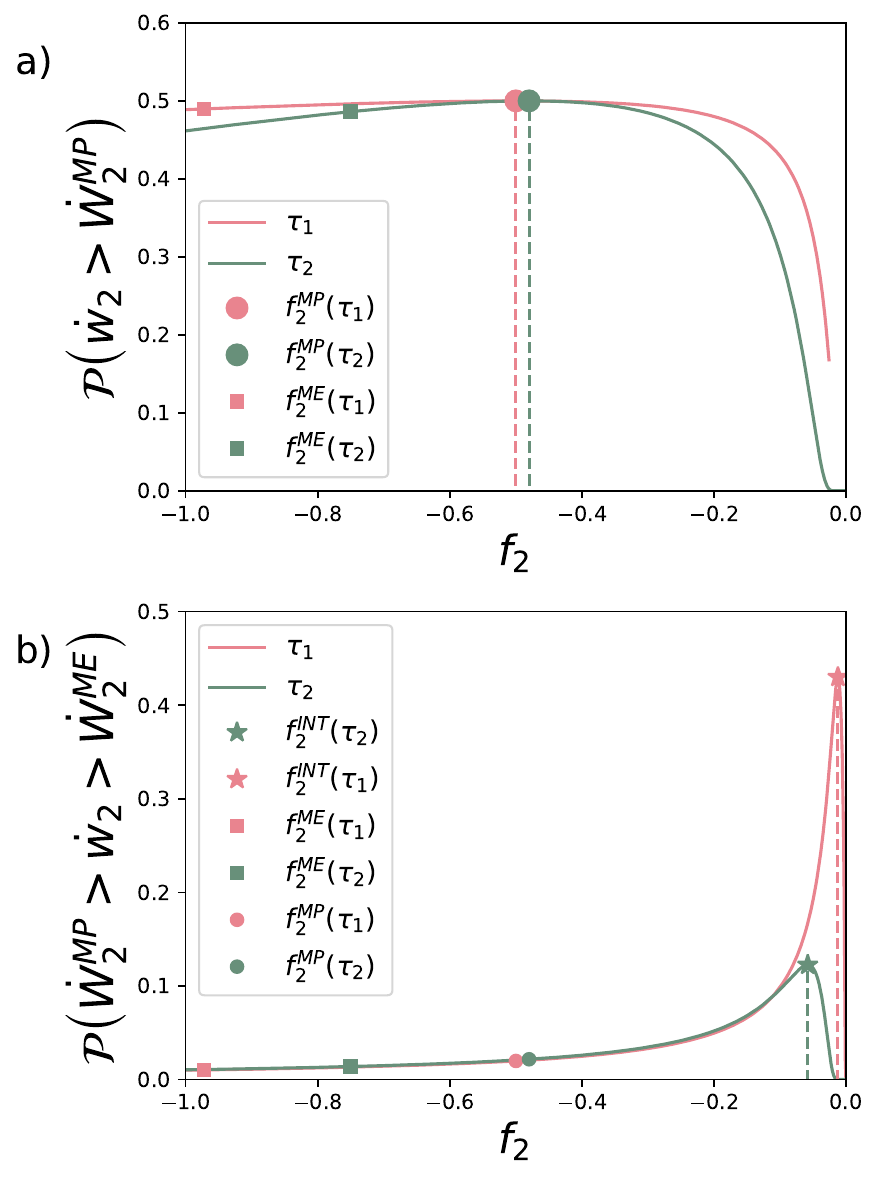}  
    \caption{Probability distribution of power ${\dot W}_2$. (a) Probability that power exceeds its maximum mean value, which is maximized at the force corresponding to the maximum average power. (b) Probability that power falls within an intermediate range, which is maximized at a distinct value $f_2^{\rm{INT}}$. For $\tau_1 = 0.1$ and $\tau_2 = 1$, the values of $f_2^{INT}$ are approximately $-0.0126$ and $-0.0583$, yielding probabilities of $43\%$ and $12\%$, respectively.}  
    \label{potentialprobability}  
\end{figure}  

\begin{equation}  
    \mathcal{P}\left(\bar{\dot w}_2>\dot W_2^{\rm MP} \right)  =
\frac{1}{2} \left( 1 + \erf\left(\frac{\dot W_2 - \dot W_2^{\rm MP}}{\sqrt{2} \sqrt{C_{22}}}\right) \right).
\end{equation}  
The above expression highlights a fundamental property of the Gaussian distribution: when $\dot W_2 = \dot W_2^{MP}$, the error function vanishes, yielding a probability of exactly 50\%. This result is expected, as it reflects the symmetry of the Gaussian distribution about its mean. Consequently, the probability of exceeding the maximum mean power is inherently bounded by $\mathcal{P}(\bar{\dot w}_2>\dot W_2 ) \leq 50\%$. In Figure~\ref{potentialprobability}(a), we illustrate the probability of power exceeding its maximum mean value as a function of $f_2$, evaluated at two distinct time intervals, $\tau_1=0.1$ and $\tau_2=1$. The highest probability, reaching 50\%, occurs at the value of $f_2$ that maximizes the mean power, as determined by Eq.~(\ref{maxpower}). Notably, for $\tau_1$, the probability remains high across a broader range of $f_2$ values, indicating that multiple values of $f_2$ can yield a significant chance of exceeding $\dot W_2^{MP}$. Conversely, for $\tau_2$, the probability distribution is sharper, indicating that as the time interval increases, the probability of surpassing the maximum power becomes more localized around specific $f_2$ values. This suggests that, at longer timescales, fewer configurations allow for fluctuations that significantly exceed the mean maximum power.

Another relevant statistical event is the probability of power falling within an intermediate range, specifically between the average power at maximum efficiency and the maximum power. Since $\bar{\dot{W}}_2^{\rm MP} > \bar{\dot{W}}_2^{\rm ME}$, we define this probability as  
\begin{eqnarray}
    \mathcal{P}\left(\dot W_2^{\rm MP} >\bar{\dot w}_2> \dot W_2^{\rm ME} \right) \nonumber\\= 
\frac{1}{2} \left( \erf\left(\frac{\dot W_2 - \dot W_2^{\rm ME}}{\sqrt{2} \sqrt{C_{22}}}\right) - \erf\left(\frac{\dot W_2 - \dot W_2^{\rm MP}}{\sqrt{2} \sqrt{C_{22}}}\right) \right).
\end{eqnarray}  
Note that, if $\dot{W}_2 \sim \dot{W}_2^{\rm MP}$, this probability decreases, while if $\dot{W}_2 \sim \dot{W}_2^{\rm ME}$, it increases, suggesting a competition between the two regimes. In Figure~\ref{potentialprobability}(b), we show the probability for the intermediary range over values of $ f_2 $ for different time intervals. Unlike the previous case, this probability is maximized at a different force value, denoted $ f_2^{\rm INT} $, which does not coincide with either $ f_2^{\rm MP} $ or $ f_2^{\rm ME} $. Notably, in both cases depicted in Figure~\ref{potentialprobability}, increasing the time interval reduces the probability. This follows from Eq.~(\ref{covariance}), where a longer interval decreases the variance, leading to fewer fluctuations and a lower probability of significant deviations from the average. We underscore that this result holds regardless of the specific protocol used, as the variance formula is protocol-independent, and the power distribution is Gaussian regardless of the protocol.

\begin{figure}
    \centering
    \includegraphics[width=8.6cm]{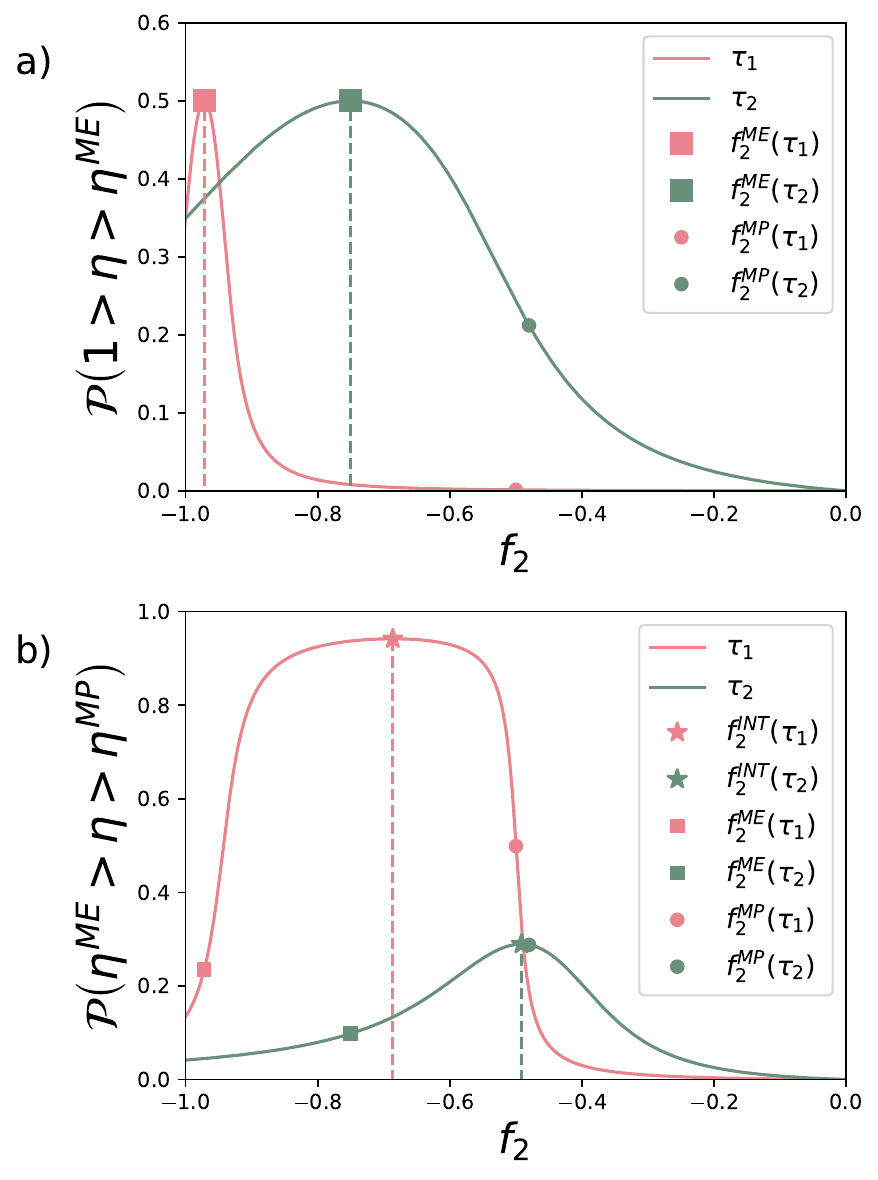}
    \caption{Probabilities for the efficiency. a) Probability to have the efficiency being greater than the efficiency at maximum mean power. For both time intervals, the maximum probability is 50\%. b) Probability to have the efficiency being greater than the mean efficiency at maximum power, but less than the mean maximum efficiency. The optimal force values maximizing this probability are $f_2^{\mathrm{INT}} = -0.686$ for $\tau_1 = 0.1$ and $f_2^{\mathrm{INT}} = -0.492$ for $\tau_2 = 1$, yielding probabilities of approximately $90\%$ and $20\%$, respectively.}
    \label{efficiencyprobability}
\end{figure}
\subsection{Efficiency optimization}
Building on our investigation of power fluctuations, we now turn to the statistical properties of the system's efficiency. The work-to-work converter operates within the well-defined efficiency range of $0 \leq \eta \leq 1$, with values exceeding $\eta > 1$ representing a change in the roles of input and output force (i.e., efficiency is redefined as $\eta_{\mathrm{new}}\to 1/\eta$). Our primary interest lies in quantifying the probability of observing efficiencies that surpass the maximum mean efficiency, expressed as $\mathcal{P}(1 > \eta > \eta^{\mathrm{ME}})$. Unlike the power distribution, the efficiency probability distribution requires numerical evaluation due to its more complex mathematical form. Figure~\ref{efficiencyprobability}(a) reveals that this probability peaks at $50\%$ when $f_2 = f_2^{\mathrm{ME}}$, mirroring the optimal condition for mean efficiency. This similarity with the power statistics, however, hides a crucial difference: while power fluctuations diminish with increasing time intervals, efficiency maintains significant probability density across a broader range of $f_2$ values. This fundamental distinction stems from the Cauchy-like nature of the efficiency distribution \cite{verley2014universal}.

Finally, we examine the probability of observing efficiencies in the intermediate range between maximum efficiency and maximum power conditions ($\eta^{\mathrm{ME}} > \eta > \eta^{\mathrm{MP}}$). The results are shown in Figure~\ref{efficiencyprobability}(b). The optimal force values maximizing this probability consistently fall between the $f_2^{\mathrm{ME}}$ and $f_2^{\mathrm{MP}}$ benchmarks identified in Section~\ref{IV}. The time dependence of these probabilities reveals an important trend: while the short-time case ($\tau_1$) shows a high probability for intermediate efficiencies, this value drops substantially for the longer interval ($\tau_2$). This $\tau$ dependence, combined with the broader $f_2$ sensitivity noted earlier, highlights how efficiency statistics differ fundamentally from their power counterparts.

Our examination of independent probabilities for both power and efficiency sets the stage for the crucial next step. Since both quantities derive from the same underlying stochastic trajectories, their statistical correlation becomes essential for a complete understanding of the system's behavior. In the following section, we will investigate how these intrinsic correlations influence the joint probability structure and what implications they hold for the converter's operational characteristics.

\section{Conditional probability of power and efficiency}
\label{VI}
\begin{figure*}[htbp]
    \centering
\includegraphics[width=0.95\textwidth]{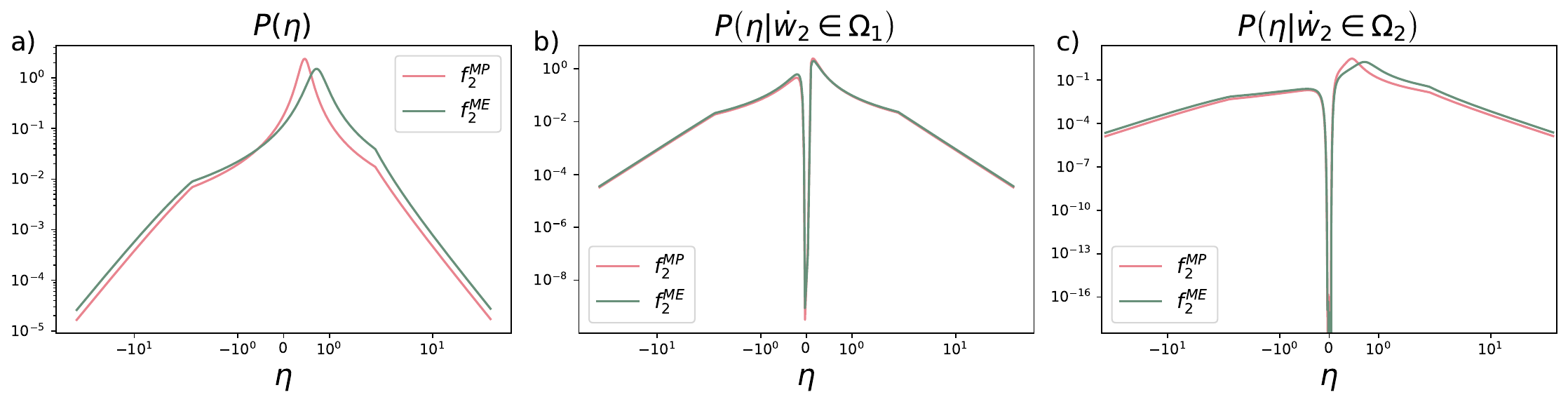}
    \caption{Probability distributions. a) Probability distribution for the efficiency. b) Conditional probability distribution for the efficiency with the condition of $\dot{w}_2 \in \Omega_1 = [\bar{\dot{W}}_2^{ME},\bar{\dot{W}}_2^{MP}]$. c) Conditional probability distribution for the efficiency with the condition of $\dot{w}_2 \in \Omega_1 = [\bar{\dot{W}}_2^{MP},\infty)$.
   All the distributions are compared for two thermodynamic force values, $f_2^{MP}$ and $f_2^{ME}$. For all plots, we choose $\tau=\tau_2=1$.}
    \label{distributions}
\end{figure*}

The joint distribution of power and efficiency enables us to investigate their conditional relationship through a rigorous statistical framework. We consider $\Omega$ as an arbitrary measurable event for the power variable, which could represent the power exceeding its maximum value ($\dot{w}_2 > \dot{W}_2^{\mathrm{MP}}$) or lying within some intermediate range ($\dot{W}_2^{\mathrm{ME}} < \dot{w}_2 < \dot{W}_2^{\mathrm{MP}}$). The conditional probability distribution for efficiency given such power events is formally expressed as
\begin{equation}
    P(\eta | \bar{\dot w}_2 \in \{\Omega\} ) =  \frac{\int_\Omega P(\eta, \bar{\dot w}_2) d\bar{\dot w}_2}{\mathcal{P}(\bar{\dot w}_2 \in \{\Omega\})},\label{conditionaldist}
\end{equation}
where the denominator $\mathcal{P}(\bar{\dot w}_2 \in \{\Omega\}) = \int_\Omega P(\bar{\dot w}_2)\,d\bar{\dot w}_2$ represents the total probability of the specified power condition occurring. This conditional formulation provides a powerful tool for examining how constrained power conditions influence the efficiency distribution, revealing statistical dependencies that are not immediately apparent from the joint distribution alone. As illustrated in Figure~\ref{distributions}, this approach allows us to systematically compare different operational regimes and quantify how power fluctuations propagate to affect the system's efficiency characteristics. 
\subsection{Performance optimization: Condition 1}
\begin{figure}[h]
    \centering
\includegraphics[width=8.6cm]{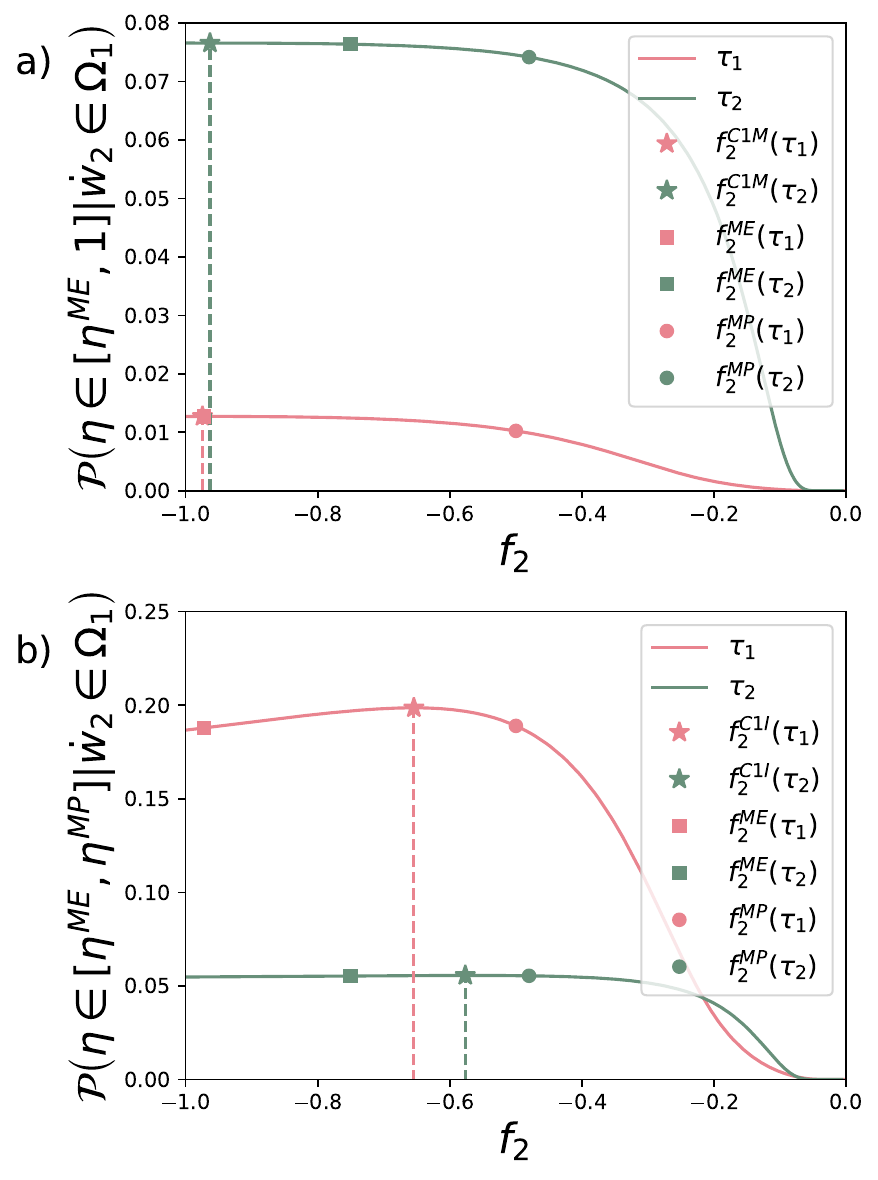}
    \caption{Conditional probability, with $\Omega_1 = [\dot W_2^{\rm ME},\dot W_2^{\rm MP}]$. a) Conditional Probability for the efficiency being greater than the average maximum efficiency. For $\tau_1 = 0.1$, the optimal force is $f_2^{\mathrm{C1M}} = -0.974$ with maximum probability of $\sim 8\%$; for $\tau_2 = 1$, the optimal force is $f_2^{\mathrm{C1M}} = -0.963$ with maximum probability of $\sim 1.2\%$. b) Conditional Probability for the efficiency being in the intermediary range. For $\tau_1 = 0.1$, the optimal force is $f_2^{\mathrm{C1I}} = -0.654$ with maximum probability of $\sim 20\%$; for $\tau_2 = 1$, the optimal force is $f_2^{\mathrm{C1I}} = -0.577$ with maximum probability of $\sim 5\%$. For all cases, the forces that maximize these probabilities differ from the reference forces.}
    \label{condition1}
\end{figure}

We begin by considering the case where power is constrained to the intermediate regime between maximum efficiency and maximum power conditions, defined as $\Omega_1 = [\dot{W}_2^{\mathrm{ME}}, \dot{W}_2^{\mathrm{MP}}]$. Using the conditional probability framework from Eq.~\eqref{conditionaldist}, we first examine the probability of efficiencies exceeding the maximum mean value ($\eta > \eta^{\mathrm{ME}}$) under this power constraint. Figure~\ref{condition1}(a) reveals several important features of this conditional probability. First, the maximum probabilities are significantly reduced compared to the unconstrained case, falling from $50\%$ to much lower values for both time intervals. The optimal force values that maximize this probability are denoted as $f_2^{\mathrm{C1M}}$, where "C1M" indicates the force that maximizes efficiency probability under condition 1. Interestingly, the $\tau_1$ case shows a broader range of $f_2$ values that maintain near-maximal probability, reflecting the stronger fluctuations at shorter time scales.

We next consider the probability of efficiencies lying in the intermediate range $\eta^{\mathrm{ME}} > \eta > \eta^{\mathrm{MP}}$ under the same power constraint. As shown in Figure~\ref{condition1}(b), the maximum probabilities occur at distinct optimal forces $f_2^{\mathrm{C1I}}$, where "C1I" represents the force that maximizes intermediate efficiency probability under condition 1. Notably, these optimal forces lie closer to the maximum-power condition than to the maximum-efficiency condition, demonstrating how the power constraint biases the system toward different operational regimes.
The systematic reduction in probabilities for both high and intermediate efficiency regimes when power is constrained to $\Omega_1$ highlights the fundamental trade-off between these quantities. While fluctuations do allow finite probabilities for favorable efficiency-power combinations, the conditional analysis reveals how power constraints limit the system's ability to simultaneously achieve high efficiency values.

\subsection{Performance optimization: Condition 2}
Extending our investigation beyond the intermediate power range, we now focus on the regime where power output exceeds its maximum mean value ($\Omega_2 = [\dot{W}_2^{\mathrm{MP}}, \infty)$). The conditional efficiency distribution for this case, presented in Figure~\ref{distributions}(c), exhibits several noteworthy features that contrast sharply with the intermediate-power results. When examining $\mathcal{P}(\eta \in [\eta^{\mathrm{ME}},1] | \dot{w}_2 \in \Omega_2)$, we observe a striking enhancement in probabilities compared to other cases. Numerical evaluation reveals probabilities consistently above $50\%$ across both time scales studied, reaching their maxima at the optimal force value $f_2^{\mathrm{MP}}$. This represents a significant increase from both the unconstrained case ($50\%$ maximum) and the intermediate-power condition. The enhancement is particularly pronounced for $\tau_1$, where probabilities approach $90\%$ for certain parameter ranges. This phenomenon is particularly interesting as it directly contradicts the classical efficiency power trade-off observed in mean values, suggesting that in regimes of strong power fluctuations, the system can achieve both high power and high efficiency by exploiting rare but favorable events where energy conversion becomes exceptionally efficient.

We now examine the probability of observing intermediate efficiency values, specifically in the range $\eta^{\mathrm{ME}} > \eta > \eta^{\mathrm{MP}}$, under the condition that power exceeds its maximum value ($\dot{w}_2 > \dot{W}_2^{\mathrm{MP}}$). Figure~\ref{condition2}(b) shows this conditional probability as a function of $f_2$ for different time intervals $\tau$. Three key observations emerge: first, the probability decreases with increasing $\tau$; second, the optimal $f_2$ values match those maximizing the non-conditional intermediate efficiency probabilities; and third, these probabilities consistently exceed their non-conditional counterparts, demonstrating that high-power conditions enhance the probability of achieving higher efficiency values. 
\subsection{Performance optimization: Comparison}
\begin{figure}
    \centering
    \includegraphics[width=8.6cm]{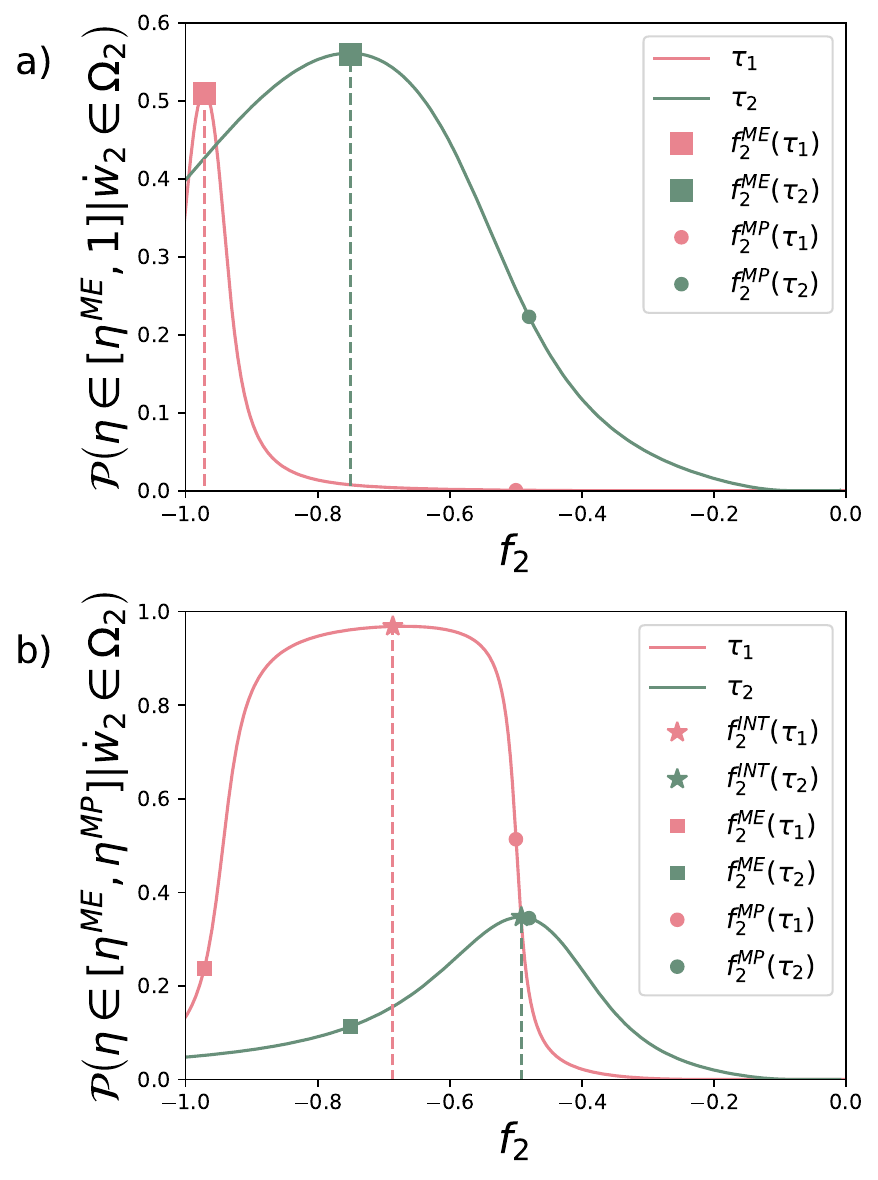}
    \caption{Conditional probability, with $\Omega_2 = [\dot W_2^{\rm MP},\infty)$. a) Conditional Probability for the efficiency being greater than the average maximum efficiency. The probabilities are higher compared with condition 1 and the case without condition, reaching $\sim 51\%$ for $\tau_1 = 0.1$. b) Conditional Probability for the efficiency being in the intermediary range. For $\tau_1 = 0.1$, the maximum probability reaches $\sim 97\%$. For all cases, the forces $f_2$ that maximize these probabilities are the same forces that maximize the probability in the unconditional efficiency case.}
    \label{condition2}
\end{figure}

The comparison between condition 1 (intermediate power) and condition 2 (high power) reveals fundamental differences in their efficiency distributions (Figure~\ref{distributions}(b)-(c)). Condition 2 produces broader distributions across $0 < \eta < 1$, with greater probability density in both high and intermediate efficiency ranges compared to condition 1's sharply peaked distribution near $\eta\sim0$. This explains why high-power trajectories are more likely to yield both maximum and intermediate efficiency values.
Physically, this reflects the stronger alignment between high-power trajectories and high-efficiency states in the Brownian particle system. When power reaches its maximum (condition 2), the corresponding trajectories are more likely to also achieve maximum efficiency, as illustrated in Figure~\ref{condition2}(a)-(b). Conversely, intermediate-power trajectories (condition 1) show weaker correlation with high-efficiency states (Figure~\ref{condition1}(a)-(b)). Table~\ref{ta2} summarizes the results by presenting the maximum probabilities obtained with and without the imposed conditions. The data clearly demonstrate that these conditions have a significant impact on the probabilities. In the example analyzed, the probability associated with $\Omega_1$ decreases, whereas the probability for $\Omega_2$ increases compared to the unconditional case.

\begin{table}[h!]
\centering
\renewcommand{\arraystretch}{1.5}
\caption{Probabilities for the efficiency with conditional constraints for the different $\tau$'s.}
\label{ta2}
\begin{tabular}{|c|c|c|c|c|}
\hline
 &  & $\cancel{\Omega}$ & $\Omega_1$ & $\Omega_2$ \\ \hline
\multirow{2}{*}{\textbf{$\tau_1$}} & $\eta \in [\eta^{\rm ME}, 1]$      & 50\%      &  1.2\%    &  51\%     \\ \cline{2-5}
                          & $\eta \in [\eta^{\rm MP}, \eta^{\rm ME}]$ & 94\%       &  19\%     &    97\%     \\ \hline
\multirow{2}{*}{\textbf{$\tau_2$}} & $\eta \in [\eta^{\rm ME}, 1]$      & 50 \%      &   7.6\%    & 55\%     \\ \cline{2-5}
                          & $\eta \in [\eta^{\rm MP}, \eta^{\rm ME}]$ & 28\%       &  5.1\%     &    35\%     \\ \hline
\end{tabular}
\end{table}

To conclude, we note that the inverse conditional probability (of power given efficiency) can be obtained through Bayes' theorem:
\begin{equation}
\mathcal{P}({\bar{\dot w}}_2 \in \Omega_w| \eta \in \Omega_\eta) = \mathcal{P}(\eta \in \Omega_\eta|{\bar{\dot w}}_2 \in \Omega_w) \frac{\mathcal{P}({\bar{\dot w}}_2 \in \Omega_w)}{\mathcal{P}(\eta \in \Omega_\eta)}. \label{cond2}
\end{equation}

This expression solely depends on the ratio of marginal probabilities, all of which have been previously computed in our analysis.

\section{Conclusion}
\label{conclusion}
In this work, we took a step forward in investigating statistical events in stochastic machines. We explored the distributions of power and efficiency and also the probabilities of efficiency and power exceeding their maximum mean values. This allowed us to examine the chance of thermodynamically advantageous outcomes arising from the stochastic nature of the Brownian system.

We have shown that, as the entire statistical framework is governed by the Onsager coefficients, our results are not sensitive to the specific protocol used, since we are working with Gaussian distributions. Using a constant force protocol to illustrate our investigation, we derived expressions for the probability of power and efficiency. These expressions enabled us to explore the probabilities of the occurrence of different events, such as $\bar{\eta} > \eta^{\textrm{MP}}$ and $\dot{W} > \bar{\dot{W}}^{\textrm{MP}}$. To better understand the trade-off between power and efficiency, we calculated the joint and conditional distributions. This allowed us to investigate the conditional probability of efficiency exceeding its maximum, as well as remaining in an intermediate range. Interestingly, we found that when power exceeds its maximum, the probability of efficiency also exceeding its maximum (or falling within the intermediate range) increases, showing the correlation between the probabilistic events. Conversely, when power is in the intermediate regime, the probability of the corresponding efficiency events drops significantly.

Our work contributes to ongoing investigations of efficiency in stochastic and quantum systems, showing how fluctuations can be understood and explored beyond standard distribution analysis. By using conditional probabilities, we can enhance or reduce the chances of specific desirable events. Our findings were derived for the specific case of position-independent forces, which result in Gaussian distributions. However, the same theoretical approach can be readily extended to more general situations involving position-dependent forces and their corresponding non-Gaussian distributions. The versatility of this approach suggests its applicability to a wide range of physical systems. For instance, in the context of thermal machines, where both position-dependent forces and dynamic temperature protocols are essential features, our framework could provide significant insights. This opens up promising directions for future research, particularly in exploring the efficiency and performance of microscopic engines.

\section*{Acknowledgement:} We thank C. E. Fiore and Luca Abrahão for useful discussions. This work was supported by the Brazilian agencies CAPES and CNPq. P.V.P acknowledges the Funda\c{c}\~ao de Amparo \`a Pesquisa do Estado do Rio de Janeiro (FAPERJ Process SEI-260003/000174/2024) and StoneLab. G.A.L. For\~ao acknowledges the financial support from Fundação de Amparo à Pesquisa do Estado de São Paulo (FAPESP) under Grant No. 2022/16192-5.  This study was financed in part by Coordena\c c\~ ao de Aperfei\c coamento de Pessoal de N\' ivel Superior - Brasil (CAPES) - Finance Code 001. F.S. Filho also acknowledges the Special Research Fund (BOF) of Hasselt University under Grant No.
R-13995.

\appendix

\section{Conditional Distributions}

The conditional distribution of the efficiency given a condition for the power,Eq.~(\ref{conditionaldist}), can be calculated analytically since the joint distribution Eq.~(\ref{jointdist}) is Gaussian.

\subsection{Condition 1}
By integrating the joint distribution over $\Omega_1$ for the power, and dividing by $\mathcal{P}(\bar{\dot w}_2 \in \Omega_1)$, we have the conditional distribution of the efficiency.
\begin{widetext}
    \begin{eqnarray}
        P(\eta|\bar{\dot w}_2 \in \Omega_1 ) = \frac{1}{2 \pi  (\eta  (C_{11} \eta +2 C_{12})+C_{22})^{3/2} \left(\text{erf}\left(\frac{\dot W_2-\dot W_2^{\rm ME}}{\sqrt{2} \sqrt{C_{22}}}\right)-\text{erf}\left(\frac{\dot W_2-\dot W_2^{\rm MP}}{\sqrt{2} \sqrt{C_{22}}}\right)\right)} \exp \left(-\frac{(\dot W_1 \eta +\dot W_2)^2}{2 (\eta  (C_{11} \eta +2 C_{12})+C_{22})}\right) \times \nonumber\\ \Bigg[2\sqrt{-\left(\left(C_{12}^2-C_{11} C_{22}\right) (\eta  (C_{11} \eta +2 C_{12})+C_{22})\right)} \Bigg(\exp \left(\frac{(\eta  (C_{11} \eta  (\dot W_2^{\rm ME}-\dot W_2)+C_{12} (\dot W_1 \eta -\dot W_2+2 \dot W_2^{\rm ME}))+C_{22} (\dot W_1 \eta +\dot W_2^{\rm ME}))^2}{2 \eta ^2 \left(C_{12}^2-C_{11} C_{22}\right) (\eta  (C_{11} \eta +2 C_{12})+C_{22})}\right) \nonumber \\ - \exp \left(\frac{(\eta  (C_{11} \eta  (\dot W_2^{\rm MP}-\dot W_2)+C_{12} (\dot W_1 \eta -\dot W_2+2 \dot W_2^{\rm MP}))+C_{22} (\dot W_1 \eta +\dot W_2^{\rm MP}))^2}{2 \eta ^2 \left(C_{12}^2-C_{11} C_{22}\right) (\eta  (C_{11} \eta +2 C_{12})+C_{22})}\right) \Bigg)+ \sqrt{2 \pi } (-C_{11} \dot W_2 \eta +C_{12} \dot W_1 \eta -C_{12} \dot W_2+C_{22} \dot W_1)\times\nonumber \\ \times\Bigg( \text{erf}\left(\frac{\eta  (C_{11} \eta  (\dot W_2^{\rm ME}-\dot W_2)+C_{12} (\dot W_1 \eta -\dot W_2+2 \dot W_2^{\rm ME}))+C_{22} (\dot W_1 \eta +\dot W_2^{\rm ME})}{\sqrt{2} \eta  \sqrt{-\left(C_{12}^2-C_{11} C_{22}\right) (\eta  (C_{11} \eta +2 C_{12})+C_{22})}}\right)\nonumber \\ -\text{erf}\left(\frac{\eta  (C_{11} \eta  (\dot W_2^{\rm MP}-\dot W_2)+C_{12} (\dot W_1 \eta -\dot W_2+2 \dot W_2^{\rm MP}))+C_{22} (\dot W_1 \eta +\dot W_2^{\rm MP})}{\sqrt{2} \eta  \sqrt{-\left(C_{12}^2-C_{11} C_{22}\right) (\eta  (C_{11} \eta +2 C_{12})+C_{22})}}\right)\Bigg) \Bigg] \nonumber \\
    \end{eqnarray}
\end{widetext}

\subsection{Condition 2}
By integrating the joint distribution over $\Omega_2$ for the power and dividing by $\mathcal{P}(\bar{\dot w}_2 \in \Omega_2)$, we have the conditional distribution for the efficiency.
\begin{widetext}
    \begin{eqnarray}
        P(\eta|\bar{\dot w}_2 \in \Omega_2 ) = \frac{1}{2 \pi  \,\text{sgn}(\eta ) (\eta  (C_{11} \eta +2 C_{12})+C_{22})^{3/2} \left(\text{erfc}\left(\frac{\dot W_2-\dot W_2^{\rm MP}}{\sqrt{2} \sqrt{C_{22}}}\right)-2\right)}\exp \left(-\frac{(\dot W_1 \eta +\dot W_2)^2}{2 (\eta  (C_{11} \eta +2 C_{12})+C_{22})}\right) \times \nonumber \\ \Bigg[-2 \text{sgn}(\eta ) \sqrt{\left(C_{11} C_{22}-C_{12}^2\right) (\eta  (C_{11} \eta +2 C_{12})+C_{22})} \exp \left(\frac{(\eta  (C_{11} \eta  (\dot W_2^{\rm MP}-\dot W_2)+C_{12} (\dot W_1 \eta -\dot W_2+2 \dot W_2^{\rm MP}))+C_{22} (\dot W_1 \eta +\dot W_2^{\rm MP}))^2}{2 \eta ^2 \left(C_{12}^2-C_{11} C_{22}\right) (\eta  (C_{11} \eta +2 C_{12})+C_{22})}\right)\nonumber \\ \sqrt{2 \pi } (-C_{11} \dot W_2 \eta +C_{12} \dot W_1 \eta -C_{12} \dot W_2+C_{22} \dot W_1) \left(\text{sgn}(\eta ) \text{erf}\left(\frac{\eta  (C_{11} \eta  (\dot W_2^{\rm MP}-\dot W_2)+C_{12} (\dot W_1 \eta -\dot W_2+2 \dot W_2^{\rm MP}))+C_{22} (\dot W_1 \eta +\dot W_2^{\rm MP})}{\sqrt{2} \eta  \sqrt{\left(C_{11} C_{22}-C_{12}^2\right) \left(C_{11} \eta ^2+2 C_{12} \eta +C_{22}\right)}}\right)-1\right) \Bigg]\nonumber \\
    \end{eqnarray}
\end{widetext}


\end{document}